\documentclass[onecolumn,prd,12pt]{revtex4}

\usepackage{graphicx}
\usepackage{dcolumn}
\usepackage{bm}
\usepackage{amsmath}
\begin{document}

\newcommand{\be}{\begin{equation}}
\newcommand{\ee}{\end{equation}}
\newcommand{\ba}{\begin{eqnarray}}
\newcommand{\ea}{\end{eqnarray}}
\def\bone{$B^{(1)}$}
\def\bone{B^{(1)}}
\def\etal{{\it et al.~}}
\def\eg{{\it e.g.~}}
\def\ie{{\it i.e.~}}
\def\DM{dark matter~}
\def\DE{dark energy~} 
\def\GC{Galactic center~} 
\def\susy{SUSY~}
\def\msun{{\rm\,M_\odot}}

\title{Prospects for detecting Dark Matter 
with neutrino telescopes 
in Intermediate Mass Black Holes scenarios}
\author{Gianfranco Bertone}
\affiliation{INFN, Sezione di Padova, Via Marzolo 8, Padova I-35131, Italy}
\begin{abstract}
Current strategies of indirect Dark Matter detection with neutrino
telescopes are based on the search for high-energy 
neutrinos from the Solar core or from the center of the Earth.
Here, we propose a new strategy based on the detection of neutrinos from
Dark Matter annihilations in {\it mini-spikes} around Intermediate Mass Black Holes. 
Neutrino fluxes, in this case, depend on the annihilation cross-section
of Dark Matter particles, whereas solar and terrestrial fluxes are
sensitive to the scattering cross-section off nucleons, a circumstance
that makes the proposed search complementary to the existing ones.
We discuss the prospects for detection with upcoming under-water and
under-ice experiments such as ANTARES and IceCube, and show that several,
up to many, sources could be detected with both experiments. A kilometer
scale telescope in the Mediterranean appears to be ideally suited for
the proposed search. 
\end{abstract}

\maketitle

\vspace{1truecm}

Current strategies of indirect Dark Matter (DM) detection with neutrino
telescopes are based on the search for high-energy 
neutrinos from the Solar core or from the center of the Earth
(for recent reviews see Refs.~\cite{Bergstrom:2000pn,Munoz:2003gx,Bertone:2004pz}). 
The idea, proposed more than 20 years ago~\cite{indirectneutrino}, is quite simple: 
if we live immersed in a halo of DM, then there is a wind of DM particles
passing through the Sun and the Earth. A fraction of all these 
particles is expected to lose enough energy, by scattering off the nuclei 
of the star or planet, 
to become gravitationally bound. Captured particles will then slowly sink at 
its center, due to subsequent scattering events, and will annihilate 
producing a steady flux of high energy neutrinos 
(see Refs.~\cite{Bergstrom:2000pn} and ~\cite{Bertone:2004pz}, and references therein). 

At equilibrium, the flux of neutrinos 
is determined by the number of DM particles captured
per unit time by the Sun or Earth, a quantity that depends on the local density
(and velocity distribution) of DM and on the scattering 
cross-section of DM particles off nucleons, i.e. the same quantities
that are probed by direct detection experiments (e.g. Ref.~\cite{Chattopadhyay:2003xi}
and references therein). Unfortunately, a large portion of the DM parameter 
space will not
be accessible to neutrino telescopes in these scenarios. In particular, only the 
most optimistic neutralino models in the Minimal Supersymmetric Standard
Model will be accessible to kilometer sized experiments (e.g. Ref.~\cite{Bergstrom:1998xh}), 
while for Kaluza-Klein DM candidates, and more specifically for the B$^{(1)}$, 
first excitation of the hypercharge gauge boson in theories with Universal
Extra Dimensions~\cite{Servant:2002aq}, 
between 0.5 and 10 events per year are anticipated in kilometer scale 
neutrino telescopes~\cite{Hooper:2002gs}. 

We propose here a new strategy for the search of DM with neutrino
telescopes, based on the detection of secondary neutrinos produced
by DM annihilations in 'mini-spikes', i.e. recently proposed 
enhancements of DM around Intermediate Mass Black Holes (IMBHs).
We argue that mini-spikes can be bright sources of neutrinos, 
and we study the prospects for detecting them with upcoming neutrino
telescopes ANTARES~\cite{Aslanides:1999vq} and 
IceCube~\cite{Ahrens:2003ix}. 

Mini-spikes result from the reaction of DM
mini-halos to the formation or growth of IMBHs. 
In fact, the 
{\it adiabatic} growth of a massive object at the center of a 
power-law distribution of DM with index $\gamma$, induces 
a redistribution of matter into a new power-law with index
$\gamma_{sp}=(9-2\gamma)/(4-\gamma)$
~\cite{peebles:1972}.
This formula is valid over a region of size $R_s \approx 0.2 r_{BH}$, 
where $r_{BH}$ is the radius of gravitational influence
of the black hole, defined implicitly as $M(<r_{BH})=M_{BH}$,
with $M(<r)$ mass of the DM  distribution within a 
sphere of radius $r$, and $M_{BH}$ mass of the Black Hole
~\cite{Merritt:2003qc}.

To make quantitative predictions, we focus on a specific IMBHs formation
scenario~\cite{Koushiappas:2003zn}, representative of a class of models 
where these objects form directly out of cold gas 
in early-forming DM halos, and are characterized by a large mass scale, 
of order $10^5 \msun$ (see also Ref.~\cite{Begelman:2006db}
and references therein). 
In this case, mini-spikes can be bright sources of DM annihilation radiation;
indirect searches with gamma-ray telescopes, and in particular 
with GLAST~\cite{glast}, expected to be launched in 2007, are particularly
promising~\cite{Bertone:2005xz}. However, GLAST will be 
sensitive to energies up to 300 GeV, which means that the 
most interesting feature of the annihilation spectrum, 
i.e. the presence of an
identical cut-off at an energy equal to the DM particle mass,
would be missed if the DM particle is heavy. Air Cherenkov 
Telescopes (ACTs) such as CANGAROO~\cite{canga}, HESS~\cite{hess}, 
MAGIC~\cite{magic} and VERITAS~\cite{veritas} could extend 
the observations to higher energies, but due to their 
narrow field-of-view, they are not optimized for full-sky 
surveys.
 
We show here that {\it neutrino telescopes can extend full-sky 
searches at higher energies}
. In case of detection, the better sensitivity 
and energy resolution of ACTs can then be used to effectively discriminate 
mini-spikes from ordinary astrophysical sources. 
\begin{figure}[t]
\includegraphics[width=0.8\textwidth,clip=true]{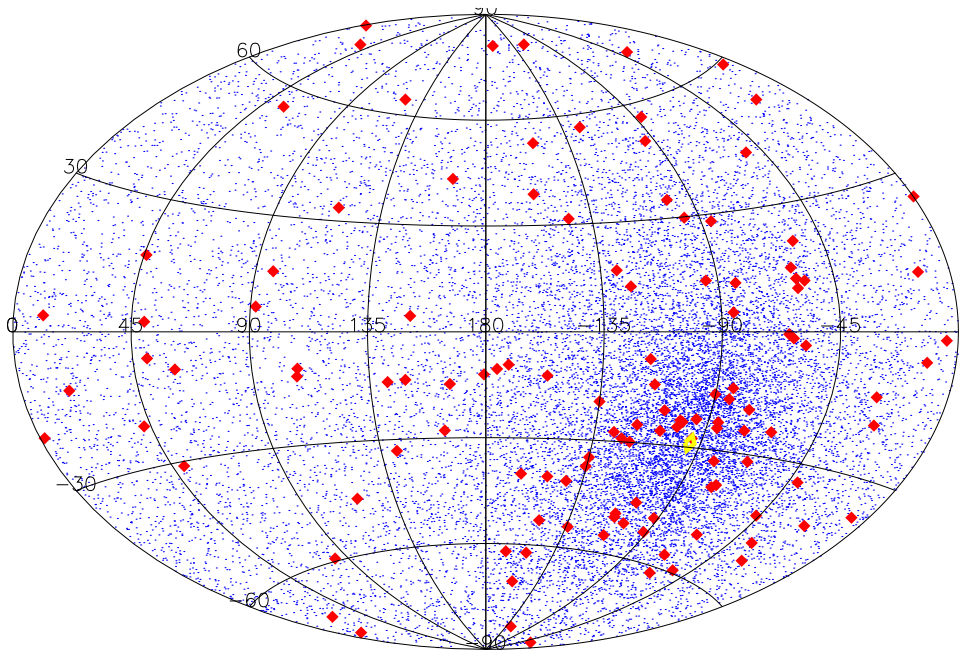}
\caption{
Sky map in equatorial coordinates showing the position 
of Intermediate Mass Black Holes in one random realization
(red diamonds), and in all 200 realizations (blue dots).
The concentration at negative declinations corresponds to 
the position of the Galactic center (black open diamond).}
\label{fig:sky} 
\end{figure}

We generate here a mock 
catalog of IMBHs, and associated mini-spikes, from the (200) realizations 
of Milky-Way like halos presented in Ref.~\cite{Bertone:2005xz},
by randomly assigning to every object
of given galactocentric distance $r_{GC}$, 2 angular
coordinates $(\theta, \phi)$, following a uniform
probability distribution on the surface of the unit sphere.
The specifics of the Monte Carlo halo evolution procedure are given in 
Ref.~\cite{Zentner:2004dq} and the method for populating black holes at
high-redshift are described in detail in Refs.~\cite{Koushiappas:2005qz,Bertone:2005xz}.
The resulting distribution is shown in Fig.~\ref{fig:sky},
where we chose to adopt equatorial coordinates, to
emphasize the different regions that will be 
accessible to Antares and Icecube.
We multiply all neutrino fluxes by a visibility factor $V_{\phi}(\delta)$,
which keeps into account that the effective area of a neutrino telescope 
at terrestrial latitude $\phi$, 
drops abruptly for sources above the horizon. The visibility factor is 
thus proportional to the fraction of time spent by a
given source of declination $\delta$ {\it below} the horizon. By means of simple
geometrical considerations we find that
\begin{equation}
V_{\phi}(\delta)=\left\{ 
\begin{array}{rr}
0, & \mbox{if } |\delta| > \pi/2-|\phi|, \delta \cdot \phi >0 \\
\arccos &\! \! \left[ \tan(\phi)  \tan(\delta) \right]/\pi, \,\,\,\,\, \mbox{if } |\delta| \leq \pi/2-|\phi| \\
1, &\mbox{if } |\delta| > \pi/2-|\phi|, \delta \cdot \phi <0
\end{array}\right.
\end{equation}
For Antares $\phi=42^\circ 50'$, while for IceCube
$\phi=-90^\circ$. In the case of IceCube the visibility factor 
reduces to 
\begin{equation}
V_{\rm{IceCube}}=\left\{ 
\begin{array}{cc}
0,&\mbox{ if } \delta < 0\\
1, &\mbox{ if } \delta >0
\end{array}\right.
\end{equation}
Mini-spikes would be copious sources of neutrinos, which 
can be produced either by the
direct annihilation of DM to neutrino pairs, or through
fragmentation and decay of secondary particles such 
as quarks, leptons and gauge bosons. 
If $N_{\nu_{\ell}}(E)$ is 
the spectrum of neutrinos $\nu_{\ell}$, of flavor $\ell=e,\mu,\tau$, per 
annihilation, the flux of neutrinos from an individual mini-spike {\it in absence
of oscillations} can be expressed as~\cite{Bertone:2005xz}
\begin{equation}
\Phi^0_{\nu_{\ell}}(E)  = \phi_0  m_{\chi,100}^{-2} (\sigma v)_{26} D_{\rm  kpc}^{-2} L_{\rm sp} N_{\nu_{\ell}}(E)
\label{eq:intrinsic}
\end{equation}
with $\phi_0 = 9 \times 10^{-10} {\rm cm}^{-2}{\rm s}^{-1}$.  
The first two factors depend on the particle physics parameters,
viz. the mass of the DM particle in units of 100 GeV 
$m_{\chi,100}$, and its annihilation
cross section in units of $10^{-26} {\rm cm}^3/{\rm s}$, $(\sigma v)_{26}$,
while the third factor accounts for the flux dilution with 
the square of the IMBH distance to the Earth in kpc, $D_{\rm  kpc}$. 
Finally, the normalization of the flux is fixed by an adimensional 
{\it luminosity factor} $L_{\rm sp}$, that depends on the specific
properties of individual spikes. In the case where the DM profile 
{\it before} the formation of the IMBH follows the commonly
adopted Navarro, Frenk and White profile~\cite{Navarro:1996he},
the final DM density $\rho(r)$ around the IMBH will be described by a 
power law $r^{-7/3}$ in a region of size $R_s$
around the IMBHs. Annihilations themselves will set an upper limit
to the DM density $\rho_{\rm max}\approx m_\chi/[(\sigma v) t]$, where 
$t$ is the time elapsed since the formation of the mini-spike, and
we denote with $R_{\rm c}$ the ``cut'' radius where $\rho(R_{\rm c})=\rho_{\rm max}$. 
With these definitions, the intrinsic luminosity factor  
in Eq.~\ref{eq:intrinsic} reads
\begin{equation}
L_{\rm sp}\equiv
\rho^{2}_{100}(R_{{\rm s}}) R_{\rm s,pc}^{14/3}
R^{-5/3}_{\rm c,mpc}
\end{equation}
where $R_{{\rm s,pc}}$ and $R_{\rm c,mpc}$ denote respectively
 $R_{{\rm s}}$ in parsecs and $R_{\rm c}$ in units of $10^{-3}$pc,
$\rho_{100}(r)$ is the density in units of $100$GeV cm$^{-3}$.
Typical values of $L_{\rm sp}$ lie in the range 0.1 -- 10~\cite{Bertone:2005xz}. 

The shape and normalization of the $\nu_{\ell}$ spectrum per annihilation
$N_{\nu_{\ell}}(E)$, depends on the nature of the DM particle.
Here we work out the
prospects for detection in two different annihilation 
scenarios, one corresponding to annihilation to $b\bar{b}$
pairs, representative of a large class of supersymmetric
models (see e.g. the discussion in~\cite{Bertin:2002ky}),
and one where DM annihilates directly to neutrino pairs,
representative of models without helicity suppression,
such as the aforementioned $B^{(1)}$ in theories with 
Universal extra-dimensions~\cite{Servant:2002aq}. 
In the case of
direct annihilation, we assume that all flavors are produced with 
equal branching ratios,
$N_{\nu_e} : N_{\nu_\mu} : N_{\nu_\tau} = 1 : 1 : 1$,
as is appropriate for Kaluza--Klein DM~\cite{Servant:2002aq}. The
neutrino spectrum per annihilation  
is in this case mono-energetic $N_{\nu_{\ell}}(E) = 2 \delta(E-m_\chi)/m_{\chi}$.
The case of annihilation to $b\bar{b}$ is more complicated,
since in this case neutrinos originate from the decay of hadrons
produced in quark fragmentation, as well as from semi-leptonic 
decay of $b$ quarks (see e.g. the discussion in ref.~\cite{Bertone:2002ms}). 
A typical fragmentation and decay chain is $b\rightarrow \pi ^{\pm} + ...$,
$ \pi ^{\pm} \rightarrow \mu + \nu_\mu $, $ \mu \rightarrow e + \nu_\mu + 
\nu_e$. Here we fold pion spectra obtained from fragmentation functions 
~\cite{Kretzer:2000yf}, with the spectra of neutrinos from 
pion decay (see e.g.~\cite{Lee:1996fp}). For semi-leptonic decay,
sub-dominant at low values of $E/m_\chi$, but inducing a hardening 
of the spectrum at high energies, we follow the treatment 
of Ref.~\cite{Jungman:1994jr}. 

We focus here on the $\nu_\mu$ flux at the ground, since neutrino telescopes 
are most sensitive to muons produced by charged-current interactions 
of $\nu_\mu$ with nuclei around the detector. However, we cannot use directly
Eq.~\ref{eq:intrinsic}, because flavor ratios are modified by neutrino 
oscillations in vacuum. For the energies ($E >$ 1GeV) and 
distances ($D_{\rm  kpc} >1$) considered here, neutrino 
oscillations will be completely averaged out ~\cite{Athar:2000yw,Crocker:1999yw}. 
The flux at the detector can in this case be expressed as
 \begin{equation}
\Phi_{\nu_{\mu}}(E)=\sum_{\ell=e,\mu,\tau} P(\nu_\ell \rightarrow \nu_\mu)
\Phi^0_{\nu_{\ell}}(E)
\end{equation}
where $P(\nu_\ell \rightarrow \nu_\mu)$ is the probability for 
a $\nu_{\ell}$ to oscillate into $\nu_{\mu}$, which
depends on the neutrino oscillation matrix elements~\cite{Eidelman:2004wy}
\begin{equation}
P(\nu_\ell \rightarrow \nu_\mu)=\sum_j |U^2_{\ell j}||U^2_{\mu j}|
\end{equation}
The analysis of solar and atmospheric neutrino oscillation, as 
well as a number of reactor and accelerator experiments, severely 
constrain the oscillation matrix elements~\cite{Eidelman:2004wy}, and 
 we adopt here 
$P(\nu_\ell \rightarrow \nu_\mu)=(0.2,0.4,0.4)$~\cite{Costantini:2004ap}.
\begin{figure}[t]
\includegraphics[width=0.8\textwidth,clip=true]{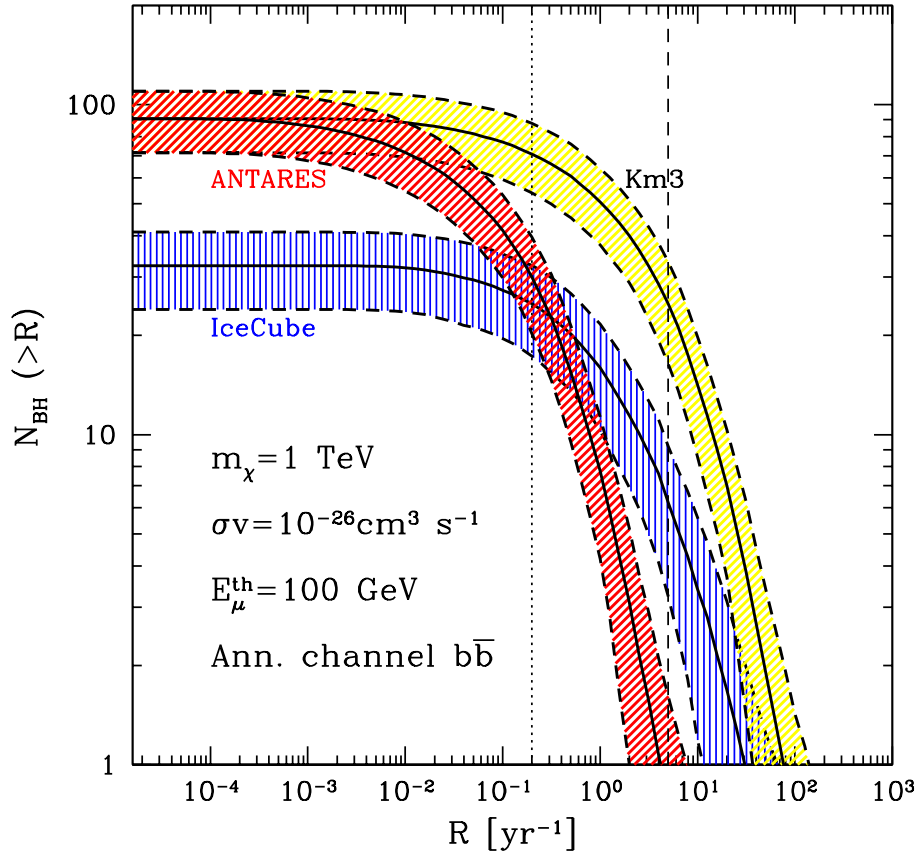}
\caption{
Number of IMBHs producing a rate $R$, or larger, in Antares, IceCube 
and Km3. Solid lines corresponds to the average over all realizations, 
while dashed line denote the 1-$\sigma$ scatter from one simulation to another.
For reference we show the rate induced by atmospheric neutrinos in Antares 
(vertical dotted) and Km3 (vertical dashed) in a $1^\circ$ cone around a 
source located at the Galactic center.}
\label{fig:1d3} 
\end{figure}
The rate of muons induced by a neutrino flux $\Phi_{\nu_{\mu}}(E)$
is given by
\begin{equation}
{\rm R} = V_\phi(\delta)\int_{E_{\mu}^{\rm thr}}^{m_\chi} dE_{\nu}
  \int_{0}^{y_{\nu}} dy\,
   A(E_{\mu})P_{\mu}(E_{\nu},y) \Phi_{\nu_{\mu}}(E_{\nu})
\label{rate}
\end{equation}
where $y_{\nu}=1-E_{\mu}^{\rm thr}/E_{\nu}$ and $E_{\mu}^{\rm{thr}}$ is the muon threshold energy of the experiment (we adopt here $E_{\mu}^{\rm{thr}}=100$ GeV). $A(E_{\mu})$ is the effective area of the detector. In our case we have used an energy-dependent $A(E_{\mu})$ for Antares~\cite{bailey:2003}, while for IceCube we assumed the indicative value $A(E_{\mu})= 1{\rm km}^2$. $P_{\mu}(E_{\nu},y)$ is the probability that a neutrino of energy 
$E_{\nu}$ interacts with a nucleon producing a muon of energy $E_{\mu} \equiv (1-y)E_{\nu}$ 
above the detector threshold energy, and can be estimated as 
 \begin{equation}
P_{\mu}(E_{\nu},y) =
 N_A\,R(E_{\mu},\, E_{\mu}^{\rm thr}) \,
\sigma(E_{\nu},y)
\end{equation}
where $R(E_{\mu},\, E_{\mu}^{\rm thr})$ is the muon range, 
{\it i.e.} the distance 
traveled by muons before their energy drops below 
$E_{\mu}^{\rm thr}$, $N_A=6.022\times 10^{23}\,g^{-1}$ 
is Avogadro's number and 
$\sigma(E_{\nu},y)\equiv d\sigma_{CC}^{\nu N}(E_{\nu},y)/dy$ is the differential
cross section for neutrino--nucleon charged--current scattering
(for further details see e.g. Ref.~\cite{Bertone:2004pz} and 
references therein).
In our treatment of high-energy neutrino-nucleon interactions 
we follow Ref.~\cite{Gandhi:1995tf}, but we make use 
of an updated set of parton distribution functions~\cite{Pumplin:2002vw}.

The results are shown in Fig.~\ref{fig:1d3}, where we plot the number of 
black holes producing a rate of events $R$, or larger, in ANTARES and IceCube,
assuming $m_\chi=1$ TeV, and $\sigma v = 10^{26}$ cm$^3$ s$^{-1}$.
For comparison, we show the rate of atmospheric neutrino events in a search 
cone of size $1^\circ$ around the source, for Antares and for a 
kilometer-scale telescope. For the atmospheric background in Antares 
we have adopted the value derived in a dedicated study~\cite{bailey:2003}, 
relative to a point source coincident with the Galactic center, while for 
kilometer scale telescopes, we have used the so-called Bartol 
flux~\cite{Agrawal:1995gk}, and a source at the same declination. 

The choice of a heavy DM particle in Fig.~\ref{fig:1d3} was motivated by
the fact that the sensitivity of neutrino telescopes increases with energy, 
due in particular to the increase of the neutrino-nucleon cross-section,
while the atmospheric background rapidly decreases with energy, a circumstance 
that leads to a better signal to noise ratio for high-energy fluxes. 
Unlike other indirect searches, {\it annihilation fluxes from 
mini-spikes around IMBHs are not very sensitive to the particle physics
parameters of the DM candidate}. In fact, the luminosity factor $L_{\rm sp}$
implicitly depends on the DM mass and cross-section, partially compensating
the $(\sigma v)/m_\chi^2$ factor in Eq.~\ref{eq:intrinsic}, so that the neutrino flux is actually 
proportional
to $(m_\chi)^{-9/7} (\sigma v)^{2/7}$, as for gamma-rays~\cite{Bertone:2005xz}.
An additional reason to focus on a heavy DM candidate is that the prospects for 
detection with gamma-ray telescopes are more promising for DM particles
lighter than 300 GeV.

As one can see from Fig.~\ref{fig:1d3}, IceCube is expected to detect a larger number of sources 
with a high event rate, with respect to Antares, and several of those are 
expected to exceed the atmospheric background. However, the sources are 
more concentrated toward 
the Galactic center, i.e. a region that cannot be seen by IceCube. The 
Antares location is more favorable, although the event rate is inevitably 
smaller due to reduced size of the experiment. A kilometer sized detector 
in the Mediterranean~\cite{km3}, combining
the large, $km^2$, area of IceCube, with the visibility function of Antares,
appears ideally suited for this type of searches.
We show in Fig.~\ref{fig:1d3} what can be achieved with a detector
with $A(E_{\mu})=1$km$^2$ at the same latitude as Antares.
Note also that the specific example in Fig.~\ref{fig:1d3}
is consistent with null searches with the Amanda-II telescope~\cite{Ackermann:2004ag}, 
although the collected data can already be used to constrain the parameter 
space of the proposed scenario.

The results for direct annihilation to $\nu_\ell \bar{\nu_\ell}$ are less
promising, unless the mass of the DM particle is just above $E_{\mu}^{\rm thr}$.
This is due the fact that in the $b \bar{b}$ channel, fragmentation and decay 
of secondary particles produce a {\it large} number of neutrinos, with energy  
much {\it smaller} than $m_\chi$. The direct annihilation thus dominates only when
most of these neutrinos are produced at energies below $E_{\mu}^{\rm thr}$.
For the parameters in Fig.~\ref{fig:1d3}, the $\nu_\ell \bar{\nu_\ell}$ 
flux is smaller by a factor $\sim 3.5$ for Antares, and $\sim 5$ for IceCube and Km3.
Note that although 
neutrinos from the Galactic center may in principle represent an interesting 
alternative, the prospects for detecting gamma-rays from the
same source are more interesting both for 
supersymmetric~\cite{Bertone:2004ag} and Kaluza-Klein DM~\cite{Bertone:2002ms,Bergstrom:2004cy}.

We recall that we derived our results in the framework of a specific IMBHs
scenario, characterized by a heavy mass scale $\sim 10^5 M_\odot$.  
Alternatively, we could have chosen to work in a different scenario,
where  IMBHs form in overdense regions at high redshift, $z \sim 20$, 
as remnants of Population III stars, and have a 
characteristic mass-scale of a few $10^2 \msun$ 
\cite{Madau:2001}. However, in this scenario the formation of IMBHs 
is not adiabatic~\cite{Bertone:2005xz}, and mini-spikes
can only  form by the subsequent growth  
of IMBHs by accretion of baryons~\cite{Zhao:2005yg,islamb:2004},
a circumstance that makes the mini-spikes more shallow, and the
whole scenario more problematic.

We have shown that upcoming neutrino telescopes 
may detect the neutrino flux from DM annihilations around IMBHs. The 
prospects for detection appear promising, especially for a 
kilometer-scale neutrino telescope in the Mediterranean. Both Antares
and IceCube should be able to detect several, up to many, IMBHs.

Unlike searches of high energy neutrinos from the Sun or the Earth,
which in most cases are only sensitive to {\it scattering} cross-section 
of DM particles off-nuclei, the proposed search is only sensitive to 
the {\it annihilation} cross section, 
a circumstance that makes the proposed 
search complementary to the existing ones. 

We stress 
the importance of a dedicated analysis that keeps into account the peculiar 
experimental details of the various neutrino telescopes, and a careful 
study of the detectability of sources in different regions of the DM 
parameter space.  

\paragraph*{Acknowledgements.} The author is grateful to 
A.~Zentner for many useful discussions and for making available the numerical
realizations of Milky-Way like halos, from which the properties and luminosity
of mini-spikes have been derived. We thank R.~Aloisio, J.~Beacom, P.~Blasi, 
T.~Bringmann, J.~Carr, N.~Fornengo, D.~Hooper, A.~Masiero, T.~Montaruli, E.~Nezri
and E.~Resconi for useful comments. This work was supported by the Helmholtz 
Association of National Research Centres.


\begin{thebibliography}{99}
\bibitem{Bergstrom:2000pn}
  L.~Bergstrom,
  Rept.\ Prog.\ Phys.\  {\bf 63} (2000) 793
\bibitem{Munoz:2003gx}
  C.~Munoz,
  Int.\ J.\ Mod.\ Phys.\ A {\bf 19} (2004) 3093
\bibitem{Bertone:2004pz}
  G.~Bertone, D.~Hooper and J.~Silk,
  Phys.\ Rept.\  {\bf 405} (2005) 279

\bibitem{indirectneutrino}
J.~Silk, K.~Olive and M.~Srednicki,
Phys.\ Rev.\ Lett.\ {\bf 55}, 257 (1985),
K.~Freese,
Phys.\ Lett.\ B {\bf 167}, 295 (1986),
T.~K.~Gaisser, G.~Steigman and S.~Tilav,
Phys.\ Rev.\ D {\bf 34}, 2206 (1986),
L.~M.~Krauss, M.~Srednicki and F.~Wilczek,
Phys.\ Rev.\ D {\bf 33}, 2079 (1986),
F.~Halzen, T.~Stelzer and M.~Kamionkowski,
Phys.\ Rev.\ D {\bf 45}, 4439 (1992),
V.~Berezinsky, A.~Bottino, J.~Ellis, N.~Fornengo, G.~Mignola and S.~Scopel,
Astropart.\ Phys.\ {\bf 5}, 333 (1996)

\bibitem{Chattopadhyay:2003xi}
  U.~Chattopadhyay, A.~Corsetti and P.~Nath,
  Phys.\ Rev.\ D {\bf 68} (2003) 035005

\bibitem{Bergstrom:1998xh}
  L.~Bergstrom, J.~Edsjo and P.~Gondolo,
  Phys.\ Rev.\ D {\bf 58} (1998) 103519
  [arXiv:hep-ph/9806293].

\bibitem{Servant:2002aq}
  G.~Servant and T.~M.~P.~Tait,
  Nucl.\ Phys.\ B {\bf 650} (2003) 391

\bibitem{Hooper:2002gs}
  D.~Hooper and G.~D.~Kribs,
  Phys.\ Rev.\ D {\bf 67} (2003) 055003


\bibitem{Aslanides:1999vq}
E.~Aslanides {\it et al.}  [ANTARES Collaboration],
arXiv:astro-ph/9907432.


\bibitem{Ahrens:2003ix}
J.~Ahrens  [IceCube Collaboration],
arXiv:astro-ph/0305196.

\bibitem{peebles:1972} Peebles, P.~J.~E.\ 1972, Astrophys.\ J.\, 
178, 371;
  J.~R.~Ipser and P.~Sikivie,
  Phys.\ Rev.\ D {\bf 35} (1987) 3695;
Hernquist, L., \& Sigurdsson, S.\ 1995, Astrophys.\ J.\  {\bf 440}, 554; 
  P.~Gondolo and J.~Silk,
  Phys.\ Rev.\ Lett.\  {\bf 83} (1999) 1719

\bibitem{Merritt:2003qc}
  D.~Merritt,
  Proc. of Carnegie Obs. Centennial Symposium 
  [arXiv:astro-ph/0301257].

\bibitem{Koushiappas:2003zn}
  S.~M.~Koushiappas, J.~S.~Bullock and A.~Dekel,
  Mon.\ Not.\ Roy.\ Astron.\ Soc.\  {\bf 354} (2004) 292

\bibitem{Begelman:2006db}
  M.~C.~Begelman, M.~Volonteri and M.~J.~Rees,
  arXiv:astro-ph/0602363.

\bibitem{glast} http://www-glast.stanford.edu/


\bibitem{Bertone:2005xz}
  G.~Bertone, A.~R.~Zentner and J.~Silk,
  Phys.\ Rev.\ D {\bf 72} (2005) 103517
  [arXiv:astro-ph/0509565].



\bibitem{canga} http://icrhp9.icrr.u-tokyo.ac.jp/index.html
\bibitem{hess} http://www.mpi-hd.mpg.de/hfm/HESS/HESS.html
\bibitem{magic} http://hegra1.mppmu.mpg.de/MAGICWeb/
\bibitem{veritas} http://veritas.sao.arizona.edu/index.html

\bibitem{Zentner:2004dq}
  A.~R.~Zentner, A.~A.~Berlind, J.~S.~Bullock, A.~V.~Kravtsov and R.~H.~Wechsler,
Astrophys.\ J.\  {\bf 624} (2005) 505

\bibitem{Koushiappas:2005qz} 
 S.~M.~Koushiappas and A.~R.~Zentner,
  
Astrophys.\ J.\  {\bf 639}, 7 (2006) 

\bibitem{Navarro:1996he}
  J.~F.~Navarro, C.~S.~Frenk and S.~D.~M.~White,
  Astrophys.\ J.\  {\bf 490} (1997) 493.

\bibitem{Bertin:2002ky}
  V.~Bertin, E.~Nezri and J.~Orloff,
  Eur.\ Phys.\ J.\ C {\bf 26} (2002) 111

\bibitem{Bertone:2002ms}
  G.~Bertone, G.~Servant and G.~Sigl,
  Phys.\ Rev.\ D {\bf 68} (2003) 044008
  [arXiv:hep-ph/0211342].

\bibitem{Kretzer:2000yf}
  S.~Kretzer,
  Phys.\ Rev.\ D {\bf 62} (2000) 054001

\bibitem{Lee:1996fp}
  S.~Lee,
  Phys.\ Rev.\ D {\bf 58} (1998) 043004

\bibitem{Jungman:1994jr}
  G.~Jungman and M.~Kamionkowski,
  Phys.\ Rev.\ D {\bf 51}, 328 (1995)
  [arXiv:hep-ph/9407351].


\bibitem{Athar:2000yw}
  H.~Athar, M.~Jezabek and O.~Yasuda,
  Phys.\ Rev.\ D {\bf 62} (2000) 103007
  [arXiv:hep-ph/0005104].

\bibitem{Crocker:1999yw}
  R.~M.~Crocker, F.~Melia and R.~R.~Volkas,
  Astrophys.\ J.\ Suppl.\  {\bf 130} (2000) 339
  [arXiv:astro-ph/9911292].

\bibitem{Eidelman:2004wy}
  See the review of B.~Kayser in S.~Eidelman {\it et al.}  [Particle Data Group],
  Phys.\ Lett.\ B {\bf 592} (2004) 1.
\bibitem{Costantini:2004ap}
  M.~L.~Costantini and F.~Vissani,
  Astropart.\ Phys.\  {\bf 23} (2005) 477
  [arXiv:astro-ph/0411761].

\bibitem{bailey:2003} D.~Bailey, PhD thesis, unpublished.

\bibitem{Gandhi:1995tf}
R.~Gandhi, C.~Quigg, M.~H.~Reno and I.~Sarcevic,
Astropart.\ Phys.\  {\bf 5} (1996) 81
[arXiv:hep-ph/9512364].

\bibitem{Pumplin:2002vw}
  J.~Pumplin, D.~R.~Stump, J.~Huston, H.~L.~Lai, P.~Nadolsky and W.~K.~Tung,
  JHEP {\bf 0207} (2002) 012


\bibitem{Agrawal:1995gk}
  V.~Agrawal, T.~K.~Gaisser, P.~Lipari and T.~Stanev,
  Phys.\ Rev.\ D {\bf 53} (1996) 1314

\bibitem{km3} 
http://www.km3net.org

\bibitem{Ackermann:2004ag}
  M.~Ackermann {\it et al.}  [The AMANDA Collaboration],
  Phys.\ Rev.\ D {\bf 71} (2005) 077102
  [arXiv:astro-ph/0412347].

\bibitem{Bertone:2004ag}
  G.~Bertone, E.~Nezri, J.~Orloff and J.~Silk,
  Phys.\ Rev.\ D {\bf 70} (2004) 063503
  [arXiv:astro-ph/0403322].

\bibitem{Bergstrom:2004cy}
  L.~Bergstrom, T.~Bringmann, M.~Eriksson and M.~Gustafsson,
  Phys.\ Rev.\ Lett.\  {\bf 94} (2005) 131301

\bibitem{Madau:2001} Madau, P., \& Rees, 
M.~J.\ 2001, Astrophys.\ J.\, 551, L27 

\bibitem{Zhao:2005yg}
  H.~S.~Zhao and J.~Silk,
  Phys.\ Rev.\ Lett.\  {\bf 95} (2005) 011301.

\bibitem{islamb:2004}
R.~Islam, J.~Taylor and J.~Silk,
 Mon.\ Not.\ Roy.\ Astron.\ Soc.\  {\bf 354} (2004) 427



\end{thebibliography}
\end{document}